\documentclass[12pt,a4paper,onecolumn]{article}

\usepackage{amsmath}
\usepackage{amssymb}
\usepackage[utf8]{inputenc}
\usepackage{graphicx}
\usepackage{euscript}
\usepackage{longtable}
\usepackage{array}
\usepackage{topcapt}
\usepackage{caption}
\usepackage[usenames]{color}
\usepackage{colortbl}
\usepackage{xcolor}
\usepackage[T1]{fontenc}
\usepackage{authblk}

\title{Study of volume and surface plasmons in small silicon-hydrogen nanoclusters by GW method}
\author[1,2]{N.L. Matsko}

\affil[1]{\footnotesize Skolkovo Institute of Science and Technology, Moscow, Russia, 143026}
\affil[2]{P.N. Lebedev Physical Institute, Russian Academy of Sciences, Leninskii prosp. 53, 119991 Moscow, Russia}

\date{}

\begin{document}
\maketitle

\begin{abstract}
Numerical calculations of surface and volume plasma excitations in silicon and silicon-hydrogen nanoclusters in the range Si$_{10}$-Si$_{60}$ and Si$_3$H$_8$-Si$_{39}$H$_{40}$ are performed. Some nanocluster structures were obtained using the evolutionary algorithm, others were taken from the database. The GW method was used to calculate the response function and self-energy of the structures under study.
The applied method shows the results consistent with the experiment (except plasmaron artifacts) and sufficient sensitivity allowing to investigate the effect of the cluster structure and size on the specific properties of plasma excitations.
In the studied silicon and silicon-hydrogen nanoclusters the surface is one of the key factors affecting the properties of the plasmons. Passivation of silicon dangling bonds on cluster surface changes frequency of plasmons and significantly decreases their damping. It makes the surface and volume plasmons to be clearly distinguishable even in small clusters.
\end{abstract}

\section{Introduction}
The prospects for the use of nanoscale components in industry are very promising. In addition to the higher density of the components, nanotechnology reveals opportunities associated with the special properties of perspective devices - the quantization of electronic excitations in nanoscale systems. Changing shape and size of the device one can change its resonance properties. Plasma oscillations in nano-objects are of great interest among these excitations. The practical application of nanoplasmonics will help to create devices with work frequency much higher than the frequency of modern electronics, optoelectronic devices with record performance and response time, etc. This requires an accurate theoretical description for the excitations in nano-objects.

Plasma excitations in the bulk could be divided into volume plasmons (collective charge oscillations inside the object volume)  and surface plasmons (collective charge oscillations  propagating along the surface of the object). Basically the frequency of volume plasmons ($\omega_p$) is noticeably higher than that of surface plasmons ($\omega_s$). For the jellium-vacuum surface $\omega_s=\omega_p/ \sqrt{2}$. In bulk silicon $\omega_p=16,9-17,3$ eV \cite{si_Wpl}-\cite{si_Wpl_Spl} and $\omega_s=10-10,4$ eV \cite{si_Wpl_Spl}, \cite{si_Spl}.

In nanoclusters the plasmon properties undergo considerable changes. Atomic structure of small nanoclusters generally differs greatly from that of the bulk structures and relatively big particles (thousands of atoms and more). The nanocluster structure varies widely with cluster size and composition. This, in turn, affects the electron screening and oscillator strength. Furthermore, surface atoms constitute a significant part of small nanoparticles. Electron screening in the surface region is usually weaker than in the internal region and is influenced by the cluster environment \cite{tiggesbaumker}-\cite{liebsch}. In addition, the decrease of the surface area increases plasmon scattering and plasmon peak width \cite{mitome}-\cite{wang}.
Summarizing the above we can say that plasmon properties of each individual nanocluster are unique and can vary a lot.

A large number of experimental works is devoted to study of plasmon properties in metal and semiconductor nanoclusters, nanowires \cite{tiggesbaumker},\cite{mitome}-\cite{nakashima}. The results show that the frequencies and damping of plasmon excitations in general increase as the sample decreases.
For silicon nanoparticles there are available experimental data on plasmon measurements for clusters down to 3-4 nm in diameter. As the particle size decreases from bulk to 3 nm, the plasmon frequency increases by 1 eV. In this range of particle sizes the volume plasmons (VP) have noticeably greater intensity than the surface plasmons (SP).
It can be noted that the experiment mainly gives average information on structure and surface of the nanoclusters in ensemble. Determination of the structure and properties for a given nanocluster is usually problematic.

Most of the theoretical works considered spherical nanoparticles with the wave function of electrons and holes in the form of spherical harmonics. Within the effective mass approximation it was shown that in spherical semiconducting nanoparticles with a diameter of 3-5 nm and greater (limits of applicability of the approximation) electron excitations basically undergo a blue-shift as particles decrease \cite{efros,brus}.
In the TDDFT approximation it was shown that for jellium spheres containing 20-198 electrons the frequency of volume plasmons increases as the "particle" radius decreases. For the surface plasmons the two competing effects influencing the value of frequency were indicated: the blue-shift due to the carrier confinement and the red-shift due to the surface "diffuseness" of the electronic charge. The dominant effect cannot be specified a priori \cite{ekardt}.

It can be noted that all mentioned theoretical works deal with simplified model objects and do not consider plasmons in the structures corresponding to the real stable nanoobjects. Thus, the reliable first-principles description of plasmon behavior in nanostructures, relations between plasmon properties and nanocluster structure are essential problems. Although DFT-based methods are general for calculating ground-state properties of solids, molecules and nanoparticles, they are not suitable for an accurate description of elementary excitations and dynamical properties. Information on the excited states of the system under study is needed. For this purpose the GW method was used in the present work for calculations of plasmons in silicon and silicon-hydrogen nanoclusters with size from 4 to 11,5 \AA.

\section{ Computational details}

Density functional calculations in the Quantum Espresso (QE) \cite{qe} code were used as a starting point for the one-iteration GWA (G$_0$W$_0$). The QE calculations were made with PBE GGA pseudopotential and a plane wave basis set having the cutoff energy of 50 Ry. Computations were performed for cubic supercell geometry with the side length of 45 Bohr.
Nontrivial geometries for Si$_3$H$_8$, Si$_4$H$_{10}$, Si$_7$H$_{14}$, Si$_7$H$_{16}$, Si$_{10}$, Si$_{10}$H$_{16}$, Si$_{10}$H$_{22}$, Si$_{14}$H$_{20}$, Si$_{16}$, Si$_{16}$H$_{22}$, Si$_{20}$H$_{26}$, Si$_{26}$H$_{30}$, Si$_{30}$H$_{34}$, Si$_{35}$H$_{36}$, Si$_{39}$H$_{40}$ clusters were obtained with the evolutionary algorithm realized in the USPEX code \cite{uspex1,uspex2} (see our previous publications \cite{ourEPL,our2}). Cluster geometries for Si$_{22}$, Si$_{30}$, Si$_{39}$, Si$_{51}$, Si$_{60}$ were taken from The Cambridge Cluster Database \cite{cambridge}. The atomic structures of considered nanoclusters were relaxed using PBE GGA functional until atomic forces became less than $10^{-4}$ Ry/\AA. 

For the GWA calculations the BerkeleyGW \cite{hyb-Lou}-\cite{bgw3} package was applied. Full frequency dependence method with real-axis formalism for the inverse dielectric matrix calculations was implemented. The dielectric matrix was inversed on a non-uniform grid of 140 frequency values. Energy cutoff for the dielectric matrix was set to 2.5 Ry. Total number of bands (valence+conduction) to sum over was set to 1000. The Coulomb interaction was cutoff on the edges of cell box.

An important quantity for the object under study is the spectral function :

\begin{equation} \label{specfunc} A_n(\omega)= \frac1\pi \frac{Im[\Sigma(\omega)]}{(\omega-\epsilon_n-Re[\Sigma(\omega)])^2+(Im[\Sigma(\omega)])^2}
\end{equation} 

where $\Sigma=GW$ is the self-energy, $\epsilon_n$ - is the quasiparticle (QP) energy of the electron state n. Photoabsorption, photoelectron spectrum, electron energy loss spectroscopy (EELS) profile etc. are, roughly speaking, proportional to $A_{tot}(\omega)=\sum_n^{occ}A_n(\omega)$ (sum over all occupied electron states).
Spectral function for the valence electron state ideally consists of QP peak accompanied by the series of equidistant plasmon satellite peaks which are located below on the energy scale. These satellite peaks are positioned at multiples of plasmon energy and have decreasing intensity \cite{steiner}-\cite{arya2}. The spectral function of the GWA reproduces one satellite for the QP peak of the valence state below Fermi energy and one for the conduction state above the Fermi energy. Thus the plasma frequency in GW calculations can be considered as the distance between the QP peak and its satellite. However, the GW spectral function has a noticeable shortcoming: distance between QP peak and its satellite is significantly overestimated in comparison with the experimental plasma frequency. The reason for this is that the plasmon satellite peak in (\ref{specfunc}) is dominated by zeros of $(\omega-\epsilon_n-Re[\Sigma(\omega)])$ (i.e. poles of the Green function) and is affected by the value of Im$[\Sigma]$ much weaker \cite{langreth},\cite{plasmaron1}-\cite{plasmaron2}.
In G$_0$W$_0$ the plasma frequency is typically overestimated by a factor 1.5-1.4 \cite{langreth}-\cite{arya2},\cite{guzzo},\cite{lischner}. Our G$_0$W$_0$ calculations of $A(\omega)$ for bulk silicon give plasma frequency equal to 23.5 eV, while the experimental value is 17,3-16,9 eV \cite{si_Wpl}-\cite{si_Wpl_Spl}.
Such a feature of the spectral function was called plasmaron, a coherent state of coupled hole-plasmon pairs \cite{plasmaron1,plasmaron2}. Now it is generally recognized as an artifact of the GW method. In contrast to spectral function, peaks in Im$[\Sigma]$ are mainly influenced by $\epsilon^{-1}$ (which is included in the expression for $W=v\cdot\epsilon^{-1}$) and are directly related to plasma excitations. Thus, an analysis of the features of the Im$[\Sigma]$ function is more convenient for determining the plasma excitation frequency. We will find plasma frequency as the distance between QP peak and appropriate peak in Im$[\Sigma]$ from G$_0$W$_0$ calculations. For bulk Si in this approach our calculations show volume plasma frequency equal to 16.9 eV.

\begin{figure}[h!]
\centering
\includegraphics[width=0.4\textwidth]{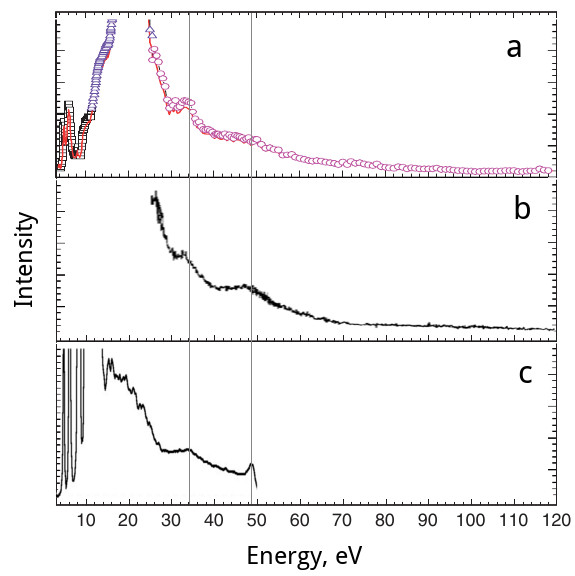}
\caption{ {\bf a} - merged data on photoabsorption and photoionization in C$_{60}$ from work \cite{kafle}: $\Box$ - photoabsorption results \cite{yasumatsu}; $\triangle$ - photoabsorption results \cite{jaensch}; $\circ$ - photoionization data \cite{kafle}. {\bf b} - photoionization cross section in C$_{60}$ from work \cite{kou}. {\bf c} - spectral density function $A_{tot}(\omega)$ inverted on the energy axis (present work GW calculations).}
\label{fig0}
\end{figure}

To test the method described above the calculations of the spectral density in the fullerene molecule C$_{60}$ were carried out. Fullerene was chosen as a semiconductor nanoscale object close in size to silicon nanoclusters considered in this paper. Moreover, there are detailed experimental data on photoabsorption and photoionization in C$_{60}$, describing plasma excitations well enough. Fig. 1a shows data on photoabsorption \cite{yasumatsu,jaensch} and photoionization  \cite{kafle} (the merged curve is given from \cite{kafle}); Fig. 1b shows photoionization data from work \cite{kou}; Fig. 1c presents spectral density function $A_{tot}(\omega)$ inverted on the energy axis (present work GW calculations).

On the spectral density graph the interval from 5.8 to 26.5 eV corresponds to valence band. In addition, the graph shows two prominent peaks at frequencies around 34 and 49 eV - plasmon satellites. The same features can be seen on the experimental curves in Fig. 1a and 1b. As mentioned above the peaks in the spectral density have frequency of 1-2 eV above the experimental ones. We can say that except for this shift, the theoretical model describes well the collective excitations in fullerene.

\section{Dependence of volume plasmon frequency on nanocluster size.}

Dependence of plasma frequency on nanocluster size can be understood qualitatively as follows. In RPA the real part of dielectric function could be written as \cite{pines}:
\begin{equation} \label{p1} \epsilon(\omega)=1-\frac{4\pi e^2}{m} \sum_{\nu}\frac{f_{0\nu}}{\omega^2-\omega^2_{\nu0}} \end{equation}
where $f_{0\nu}$ is an oscillator strength for the electron transition from the ground state to the excited state $\nu$, $\omega_{\nu0}$ - frequency of the excitation. For plasma oscillations (with frequency $\omega_p$) we can write $\epsilon(\omega_p)=0$. Assuming that plasma frequency is large compared to the characteristic interband transitions (that is valid for nanoscaled silicon clusters and larger ones), we can expand $\epsilon(\omega)$ in powers of the small parameter $\omega^2_{\nu0}/\omega_p^2$. Using the relation $\sum_{\nu}f_{0\nu}=N_{val}$, where N$_{val}$ is the number of valence electrons, we can write:
\begin{equation} \label{p2} 1-\frac{4\pi e^2}{m}\frac{N_{val}}{\omega^2_p}-\frac{4\pi e^2}{m} \sum_{\nu}\frac{f_{0\nu}\cdot\omega^2_{\nu0}}{\omega_p^4} =0 \end{equation}
Denoting $\frac{4\pi e^2}{m}N_{val}=\tilde\omega_p$ (plasma frequency for electron gas) and assuming that  $\omega_{\nu0}\approx \omega_g$ ($\omega_g$ is the HOMO-LUMO gap) we can write plasma frequency of the nanocluster as:
\begin{equation} \label{p3} \omega^2_p \simeq\ \frac12\tilde\omega_p\sqrt{ {\tilde\omega_p}^2+ 4\omega_g^2}+\frac12 \tilde\omega_p^2
\end{equation}

The dependence of the $\omega_g$ on cluster size $d$ can be approximated as the $1/d^2$  \cite{efros,brus}. Taking this and (\ref{p3}) into account it could be seen, that for small nanoclusters the plasma oscillation frequency is proportional to $1/d$. For the large clusters, when $\omega_g$ tends to the bulk value, plasmon frequency in (\ref{p3}) also tends to the bulk value.

\begin{figure}[h!]
\centering
\includegraphics[width=0.6\textwidth]{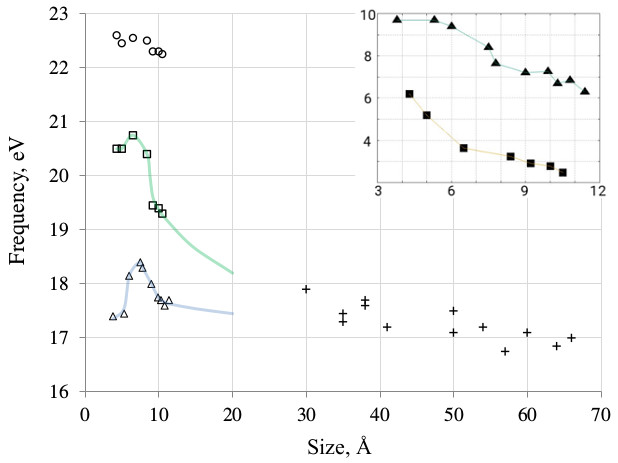}
\caption{Dependence of the volume plasmon frequency on nanocluster size. $\circ$ - frequencies from $A(\omega)$ for Si$_n$ clusters, $\Box$ - frequencies from Im$[\Sigma]$ for Si$_n$ clusters, $\triangle$ - frequencies from Im$[\Sigma]$ for Si$_k$H$_m$ clusters, $+$ - experimental frequencies from \cite{mitome,nienhaus}. The inset: filled squares and triangles - HOMO-LUMO gaps for Si and Si-H nanoclusters respectively (the axis units are the same).}
\label{fig1}
\end{figure}

Fig. 2 shows the results of GW calculations of the volume plasmon frequencies in silicon and silicon-hydrogen nanoclusters depending on particle size as well as  experimental values. Empty circles - plasmon frequencies from spectral density $A(\omega)$ calculations for silicon nanoclusters Si$_{10}$, Si$_{16}$, Si$_{22}$, Si$_{30}$, Si$_{39}$, Si$_{51}$, Si$_{60}$. Empty squares and triangles - plasmon frequencies from Im$[\Sigma]$ calculations in mentioned Si$_n$ nanoclusters and Si$_k$H$_m$ nanoclusters Si$_4$H$_{10}$, Si$_7$H$_{14}$, Si$_{10}$H$_{16}$, Si$_{14}$H$_{20}$, Si$_{16}$H$_{22}$, Si$_{20}$H$_{26}$, Si$_{26}$H$_{30}$, Si$_{30}$H$_{34}$, Si$_{35}$H$_{36}$, Si$_{39}$H$_{40}$ respectively. Crosses - experimental frequencies from \cite{mitome,nienhaus}.
The size of calculated nanoclusters was determined as the diameter of a spherical particle of the same volume.
The inset in the upper right corner shows HOMO-LUMO gaps for the studied Si and Si-H nanoclusters (filled squares and triangles respectively).

As can be seen from Fig. 2, plasma frequencies in the studied structures decrease as the cluster diameter increases for the particles with size of $d=7-9$ \AA\ and more. Si$_n$ clusters plasmon frequencies, obtained from Im$[\Sigma]$ calculations (boxes), are 2-3 eV lower than the ones from $A(\omega)$ calculations (circles). Passivation of dangling bonds in silicon clusters by hydrogen leads to a decrease in the frequency by about 2 eV (triangles).
Experimental data on plasmon frequencies are mainly available for silicon nanoclusters of 30 \AA\ and more, which is much larger than the calculated structures. However, the given experimental frequencies can help to estimate the accuracy of the method; when extending size-frequency dependence from the calculated structures to the region of larger clusters, this dependence should connect with experimental data smoothly and monotonously. Extrapolating the computed plasma frequencies of spectral function calculations for the Si$_n$ clusters (circles) we get considerably overestimated values, like in the case of bulk silicon. In turn, for $d>7-9$ \AA\ the obtained plasmon frequencies for Si and Si-H clusters from Im$[\Sigma]$ calculations together with the experimental data show dependence close to $1/d$ from formula (\ref{p3}). For large diameters the plasmon frequencies in Si and Si-H nanoclusters tend to the common bulk limit of plasma frequency in crystal silicon.

In the specified area around 7\AA\ (Si$_{22}$ cluster) and around 9\AA\ (Si$_{14}$H$_{20}$), the plasmon frequencies for Si and Si-H clusters have the maximum. With a further decrease of cluster size the frequency also begins to decrease. From the graph of HOMO-LUMO gaps for Si and Si-H clusters (inset in Fig. 2) it is clear that these maxima are not related to the electron spectrum peculiarities. In the work \cite{ekardt}  the possibility of the plasmon frequency red-shift when particle size decreases was explained by the surface "diffuseness" of the electronic charge. In our calculations Si clusters with dangling surface bonds and Si-H clusters with passivated ones have different surface "diffuseness". However, plasmon frequency dependencies for these two groups of nanoclusters demonstrate almost the same features. It looks like a contradiction with the concept of the impact of the surface charge "diffuseness". We think that such change in the plasmon frequency behavior occurs because the nanocluster size becomes comparable to the area of the electronic charge localization in the cluster. Thus, further reduction of the cluster size significantly damps the charge fluctuations.

For the small Si$_n$ clusters like Si$_{16}$ and smaller the spectral density in addition to the main QP and plasmon peaks acquires nonessential small peaks. They can be associated with the influence of dangling bonds that leads to electron states splitting. The role of the surface dangling bonds increases with nanocluster size decreasing. This affects the plasmons, since the plasma excitations can couple with interband transitions of comparable frequency \cite{wang,wilson}. Moreover, with the decrease of the surface area the scattering of plasmons and its coupling with each other increases. As a result, plasma excitations become shifted and highly damped. So the concepts of single-pair excitation or collective mode lose its meaning with the further reduction of the particle size. Passivation of dangling bonds changes situation dramatically. Even in small hydrated nanoclusters as Si$_3$H$_{8}$ and Si$_4$H$_{10}$ the plasmon peaks are clearly identified and well separated. Only with further size reduction to clusters Si$_2$H$_{6}$ and SiH$_4$ the plasmon peaks disappear, turning into a broad tail at frequencies below the QP peaks.

\section{Influence of dangling bonds passivation on volume and surface plasmons.}

In the EELS experiments for silicon nanoparticles the SP manifest themselves as peaks at frequencies 2-9 eV below the VP \cite{mitome,nienhaus}.  In the work \cite{mitome} for the 35 \AA\ silicon nanocluster one can observe several peaks (corresponding to the surface plasmons) with frequencies that are 2-5 eV lower than the dominant VP peak. As cluster size increases the distance between SP and VP peaks increases as well.  For the 100 \AA\ silicon nanoclusters and larger the SP represent a broad peak separated from the VP peak by about 7-8 eV.  Another size-related feature is relative intensity of plasmons. The change of nanocluster size leads to the change of surface area to volume ratio. For the small clusters where fraction of surface atoms is high the ratio of the SP to VP intensity becomes larger, but for clusters down to 35 \AA\ this ratio is still much less than 1 \cite{mitome,nienhaus}.

The SP properties in the examined Si$_n$ nanoclusters differ from those in Si$_k$H$_m$ principally. This distinction is illustrated by the example of Si$_{30}$ and Si$_{30}$H$_{34}$ in Fig. 3. Fig. 3a and 3b present $A(\omega)$, $Im(\Sigma)$, $G^{-1}(\omega)$ functions for the bottom valence electron state and $A_{tot}(\omega)$ for the Si$_{30}$ nanocluster respectively. Only QP and plasmaron parts are prominent in the spectral functions $A(\omega)$ for Si$_n$ clusters. For the bottom valence state of Si$_{30}$ the QP and plasmaron peaks have frequencies of -16.3 eV and -38.7 eV respectively. Collective excitations in Im$(\Sigma)$ for Si$_n$ clusters are basically represented by several peaks located at intervals of $\sim$ 2-3 eV. It cannot be separated with confidence just in appearance which peak refers to the surface and which to the volume plasmons. The Im$(\Sigma)$ graph in Fig. 3a shows four peaks corresponding to the collective excitations (indicated by arrows). The peaks are located at frequencies from -36 to -26 eV and spaced at approximately the same distances. The total spectral density $A_{tot}(\omega)$ for the Si$_{30}$ cluster (and for other studied Si$_n$) shows a broad tail at frequencies below QP peaks with no prominent peaks. The tail corresponds to collective excitations and rapidly vanishes at frequencies exceeding value of QP frequency minus VP frequency.

\begin{figure}[h!]
\centering
\includegraphics[width=0.65\textwidth]{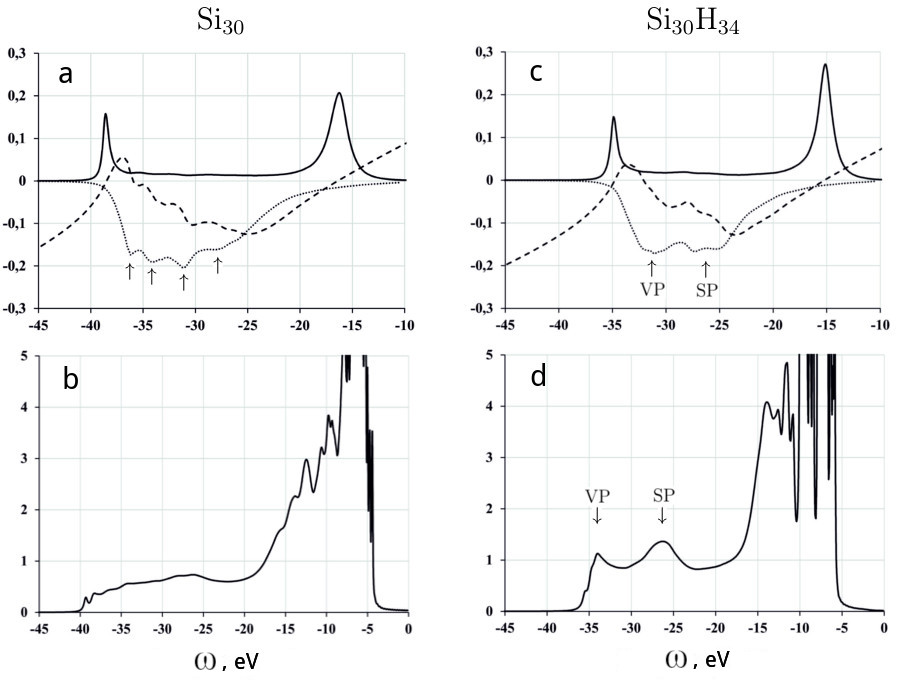}
\caption{ $A(\omega)$, $Im(\Sigma)$, $G^{-1}(\omega)$ (3a and 3c) and total $A_{tot}(\omega)$ (3b and 3d) for the Si$_{30}$ and Si$_{30}$H$_{34}$ nanoclusters. Arrows indicate the peaks corresponding to the volume and surface plasmons.}
\label{fig2}
\end{figure}

When adding hydrogen to silicon nanoclusters, the H atoms sit on the particle surface, passivating Si dangling bonds. Increasing the number of hydrogen atoms we get a cluster with a silicon core covered by H atoms. When all dangling bonds are passivated the further addition of hydrogen leads to penetration of H atoms into the structure and to the loosening of the cluster. Thus the compactness of the structure depends on the degree of passivation. The above process affects plasma excitation.
Fig. 3c presents $A(\omega)$, $Im(\Sigma)$, $G^{-1}(\omega)$ functions for the Si$_{30}$H$_{34}$ nanocluster (Si$_{30}$ nanocluster with fully passivated surface). It could be seen that plasma excitations become red-shifted with passivation. This occurs because of enhancing of the electron screening at the surface region with the passivation of the dangling bonds. Another noticeable difference from the Si$_{30}$ is that the surface and volume plasmon peaks become clearly separated from each other. The distance between surface and volume peaks increases up to 7 eV. Intensity of the peaks grows. These changes are most evident from the A$_{tot}$ graph (Fig. 3d): the long tail in the frequency range corresponding to collective modes converts to the two peaks related to the SP and VP. The mentioned influence of the hydrogen passivation on the VP and SP segregation and intensity increase is inherent to all studied nanoclusters. 

\begin{figure}[h!]
\centering
\includegraphics[width=0.7\textwidth]{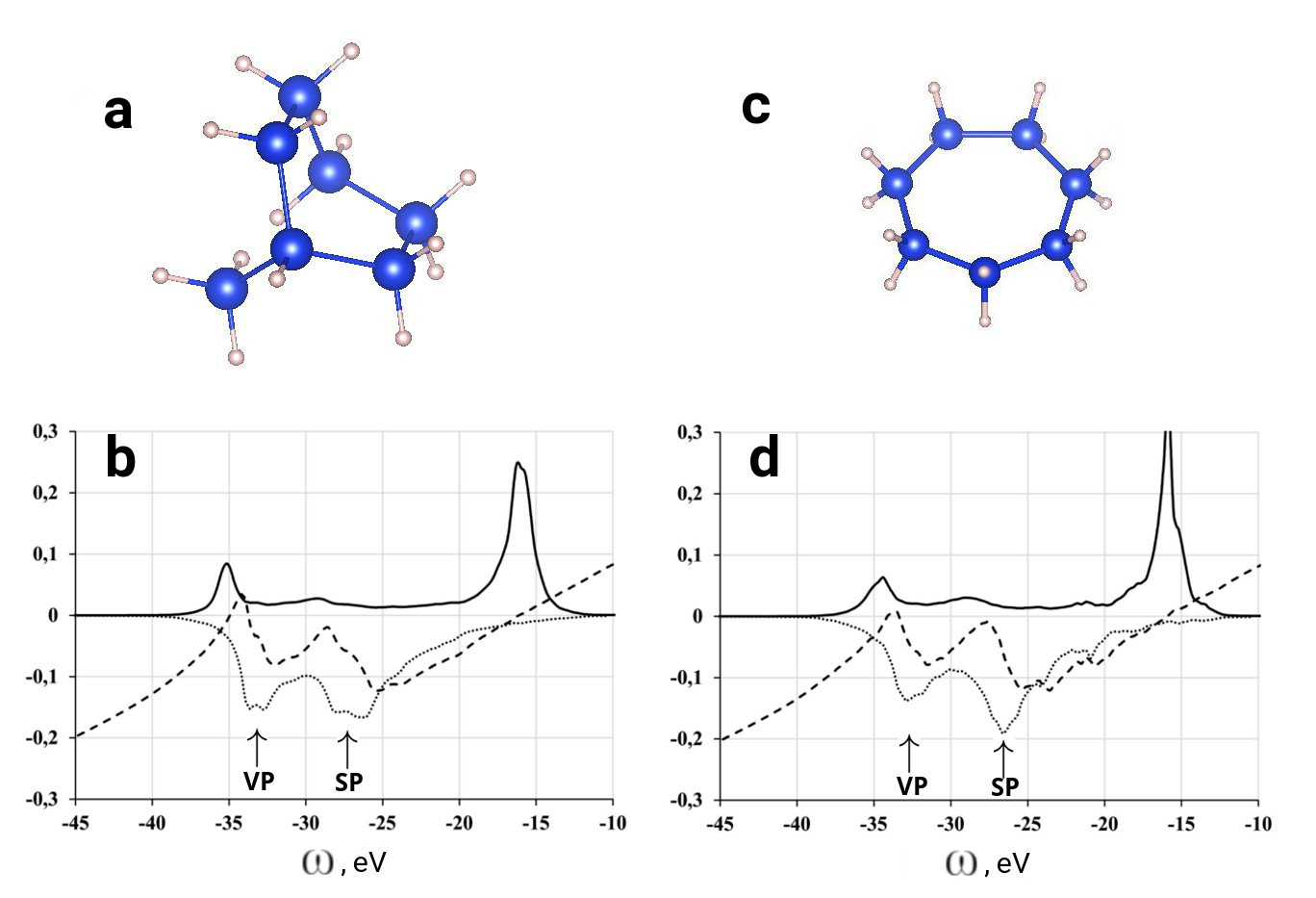}
\caption{ Structures of compact and ring Si$_7$H$_{14}$ isomers (4a and 4c).  Functions $A(\omega)$, $Im(\Sigma)$, $G^{-1}(\omega)$ for the bottom valence electron in the compact and ring isomers (4b and 4d). Arrows indicate the peaks in $Im(\Sigma)$ corresponding to the volume and surface plasmons.}
\label{fig3}
\end{figure}

Fig. 4 illustrates the influence of the nanocluster structure on plasmon properties. The upper part of Fig. 4 shows the structures for the two isomers with formula Si$_7$H$_{14}$. Isomer in Fig. 4a is compact, isomer in Fig. 4c has ring structure. Figs. 4b and 4d present relevant functions $A(\omega)$, $Im(\Sigma)$, $G^{-1}(\omega)$ for the bottom valence electron in these isomers. The ring isomer demonstrates increased intensity of the surface plasmon and slightly reduced intensity for the volume plasmon comparing to compact isomer. This can be explained if we assume that in ring isomer the surface is a smooth closed circle. Thus for the surface plasmons which propagate along the circle the scattering is lower than for plasmons which propagate along the broken surface of the compact isomer.

Relative intensity of VP and SP peaks varies  throughout the cluster set Si$_3$H$_8$-Si$_{39}$H$_{40}$. In Si$_3$H$_8$ the SP peak intensity is about 4 times more than of VP peak. In clusters Si$_{39}$H$_{40}$ and Si$_{35}$H$_{36}$ the SP and VP intensities become comparable.

\section{Conclusions}

The ratio of surface area to volume largely determines the nanocluster properties, in particular the properties of plasma oscillations. The plasmon properties are mainly similar to the bulk for particles where fraction of inner atoms is much bigger than the surface ones. VP have noticeably higher intensity than the SP, frequencies of plasmons are only slightly different from the bulk. As nanocluster size decreases the contribution of surface atoms increasingly affects the plasmon properties.
In the studied silicon nanoclusters the surface plays a key role. The surface dangling bonds in Si$_n$ clusters noticeably increase the damping of plasma excitations, coupling of SP and VP with each other and with interband transitions. Therefore, in small pure silicon clusters like Si$_{16}$ and less the concepts of "single-pair" and "collective" excitation lose their meaning, corresponding peaks become strongly blurred and intertwined. In larger Si$_n$ clusters the collective excitations are basically represented by several peaks. It cannot be separated just in appearance which peak refers to VP and which to SP.
Passivation of dangling bonds and formation of a surface layer of hydrogen atoms in Si$_k$H$_m$ nanoclusters increases intensity of VP and SP peaks, which become clearly defined and well separated from each other. This feature remains the same even for small nanoclusters like Si$_3$H$_8$.
Frequencies of volume plasmons in the studied Si$_k$H$_m$ nanoclusters are lower by about 2 eV than in Si$_k$ nanoclusters.
VP frequencies have the maxima for the clusters with diameters in the region of 7-9 \AA\ . With the increase of size of the studied nanoclusters the VP frequencies decrease tending to the bulk silicon value. The difference between plasma frequencies in Si and Si-H goes to zero. 

In the largest calculated clusters Si$_{39}$H$_{40}$ and Si$_{35}$H$_{36}$ the VP and SP have comparable intensities. When the cluster size decreases and the fraction of surface atoms increases, the intensity of SP becomes noticeably higher than that of VP. In Si$_3$H$_8$ the SP intensity is 4 times more than for VP. For isomer structures with the same formula the structure that have sufficiently larger surface to volume ratio shows higher SP intensity, while the VP intensity changes weakly.

The method used allows to calculate plasma excitations in nanoclusters containing about a hundred atoms with acceptable accuracy. Computations of clusters containing up to a thousand atoms are also of great interest, since such objects can be treated as components of plasmonic devices. In the future we plan to investigate the applicability of the model dielectric functions and the plasmon-pole models for the plasmon calculations in nanoobjects. Such approximations will significantly reduce the complexity of the calculations and will allow to study the system of larger sizes.

\section{Acknowledgements}

Calculations were carried out on the Joint Supercomputer Center of the Russian Academy of Sciences (JSCC RAS). The work was supported by the Russian Fund for Basic Research (grants 18-32-00991) and the programs of RAS. The author thanks Professor Artem R. Oganov for useful discussions.

\bibliography{thesis}

\begin{thebibliography}{99}
\bibitem{si_Wpl} H. Dinigen, Z. Phys. 180, {\bf105} (1964).
\bibitem{si_Wpl2} H.R. Philipp, H. Ehrenreich, Semiconductors and Semimetals, Volume 3, (1967).
\bibitem{si_Wpl_Spl} J. E. Rowe, G. Margaritondo, and S. B. Christman, Phys. Rev. B {\bf15}, 2195 (1977).
\bibitem{si_Spl} H. Raether, Z. Phys. {\bf171}, 436 (1966).
\bibitem{tiggesbaumker} J. Tiggesbaumker, L. Koller, K. H. Meiwes-Broer, and A. Liebsch, Phys.Rev. A {\bf48}, R1749 (1993).
\bibitem{ekardt} W. Ekardt, Phys.Rev.B {\bf31}, 6360 (1985).
\bibitem{liebsch} A. Liebsch, Phys.Rev. B {\bf48}, 11317 (1993).
\bibitem{mitome} M. Mitome, Y. Yamazaki, H. Takagi, and T. Nakagiri. Journal of Applied Physics {\bf72}, 812 (1992).
\bibitem{nienhaus} H. Nienhaus, V. Kravets, S. Koutouzov, C. Meier, and A. Lorke et al. J Vac. Sci. Technol. B {\bf24}, 1156 (2006).
\bibitem{wang} J. Wang, X. An, Q. Li, and R. F. Egerton, Appl. Phys. Lett. {\bf86}, 201911 (2005).
\bibitem{ouyang} F. Ouyang, P. E. Batson, and M. Isaacson, Phys. Rev. B {\bf46}, 15421 (1992).
\bibitem{nakashima} P. N. H. Nakashima, T. Tsuzuki, and A. W. S. Johnson, J. Appl. Phys. {\bf85}, 1556 (1999).
\bibitem{efros} Al.L. Efros and A.L. Efros, Sov. Phys. Semicond. {\bf16}, 772 (1982).
\bibitem{brus} L.E. Brus, J. Chem. Phys. {\bf80}, 4403 (1984).
\bibitem{qe} P. Giannozzi et al. J.Phys.:Condens.Matter {\bf21}, 395502 (2009).
\bibitem{uspex1} A.R. Oganov and C.W. Glass, J. Chem. Phys. {\bf124}, 244704 (2006).
\bibitem{uspex2} A.R. Oganov, A.O. Lyakhov and M. Valle, Acc. Chem. Res. {\bf44}, 227-237 (2011).
\bibitem{ourEPL} V.S. Baturin, S.V. Lepeshkin, N.L. Matsko, A.R. Oganov and Yu. A. Uspenskii. EPL {\bf106}, 37002 (2014).
\bibitem{our2} V.S. Baturin et all. Journal of Physics: Conference Series {\bf510}, 012032 (2014).
\bibitem{cambridge} http://www-wales.ch.cam.ac.uk/~wales/CCD/Si.html
\bibitem{hyb-Lou} M.S. Hybertsen and S.G. Louie, Phys. Rev. B {\bf34}, 5390 (1986).
\bibitem{bgw2} M. Rohlfing and S.G. Louie, Phys. Rev. B {\bf62}, 4927 (2000).
\bibitem{bgw3} J. Deslippe, G. Samsonidze, D.A. Strubbe, M. Jain, M.L. Cohen, and S.G. Louie, Comput. Phys. Commun. 183, 1269 (2012).
\bibitem{steiner} P. Steiner, H. Hochst, and S. Hufner, in Photoemission in Solids II, Topics in Applied Physics 27, edited by L. Ley and M. Cardona, Springer-Verlag, Heidelberg (1979).
\bibitem{langreth} D. C. Langreth, Phys. Rev. B {\bf1}, 471 (1970).
\bibitem{arya1} F. Aryasetiawan, L. Hedin, and K. Karlsson, Phys. Rev. Lett. {\bf77}, 2268 (1996).
\bibitem{arya2} B. Holm and F. Aryasetiawan, Phys. Rev. B {\bf62}, 4858 (2000).
\bibitem{plasmaron1} L. Hedin, B.I. Lundqvist and S. Lundqvist. Solid State Communications 5(4), p. 237-239 (1967).
\bibitem{guzzo} M. Guzzo, G. Lani, F. Sottile, P. Romaniello, M. Gatti, J. J. Kas, J. J. Rehr, M. G. Silly, F. Sirotti, and L. Reining, Phys. Rev. Lett. {\bf107}, 166401 (2011).
\bibitem{plasmaron2} L. Hedin, B.I. Lundqvist, and S. Lundqvist, J. Res. Nat. Bur. Stand.  {\bf74A}, No.3, (1970).
\bibitem{lischner} J. Lischner, D. Vigil-Fowler, and Steven G. Louie, Phys. Rev. Lett. {\bf110}, 146801 (2013).
\bibitem{kafle} B.P. Kafle, H. Katayanagi, M. Prodhan, H. Yagi, C. Huang and K. Mitsuke, J. Phys. Soc. Japan. {\bf77}, 014302 (2008).
\bibitem{yasumatsu} H. Yasumatsu, T. Kondow, H. Kitagawa, K. Tabayashi, and K. Shobatake, J. Chem. Phys. {\bf104}, 899 (1996).
\bibitem{jaensch} R. Jaensch and W. Kamke, Mol. Mater. {\bf13}, 143 (2000).
\bibitem{kou} J. Kou, T. Mori, S.V.K. Kumar, Y. Haruyama, Y. Kubozono, and K.Mitsuke, J. Chem. Phys. {\bf120}, 6005 (2004).
\bibitem{pines} D. Pines, "Elementary excitations in solids", Benjamin, New York-Amsterdam (1963).
\bibitem{wilson} C. B. Wilson, Proc. Phys. Soc. (London) {\bf76}, 481 (1960).




\end{thebibliography}
\bibliographystyle{gost705}

\end{document}